\documentclass[conference]{IEEEtran}
\usepackage{amsmath}
\usepackage{mathrsfs}
\usepackage{latexsym}
\usepackage{graphicx}
\usepackage{bbding}
\usepackage{indentfirst}
\usepackage{algorithm}
\usepackage{algorithmic}

\IEEEoverridecommandlockouts

\begin{document}
\title{Distributed Spectrum-Aware Clustering in Cognitive Radio Sensor Networks}
\author{\authorblockN{Huazi~Zhang$^{1}$, Zhaoyang~Zhang$^{1, \dag}$, Huaiyu Dai$^{2, \ddag}$, Rui~Yin$^{1}$, Xiaoming~Chen$^{1}$
\\1. Department of Information Science and Electronic Engineering, Zhejiang University, China.
\\2. Department of Electrical and Computer Engineering, North Carolina State University, USA.
\\Email: $^{\dag}$ning\_ming@zju.edu.cn; $^{\ddag}$huaiyu\_dai@ncsu.edu
%\thanks{}
}} \maketitle

\begin{abstract}
A novel Distributed Spectrum-Aware Clustering (DSAC) scheme is
proposed in the context of Cognitive Radio Sensor Networks (CRSN).
DSAC aims at forming energy efficient clusters in a self-organized
fashion while restricting interference to Primary User (PU) systems.
The spectrum-aware clustered structure is presented where the
communications consist of intra-cluster aggregation and
inter-cluster relaying. In order to save communication power, the
optimal number of clusters is derived and the idea of groupwise
constrained clustering is introduced to minimize intra-cluster
distance under spectrum-aware constraint. In terms of practical
implementation, DSAC demonstrates preferable scalability and
stability because of its low complexity and quick convergence under
dynamic PU activity. Finally, simulation results are given to
validate the proposed scheme.
\end{abstract}

\section{Introduction}
Characterized by large-scale and overlaid deployments, the emerging
Cognitive Radio Sensor Networks (CRSN) has attracted
global attention recently (see \cite{CRSN,CWSN,CWSNsurvey,Zhang-2011INFOCOM-CS,CRWSN,Zhang-2011ICC-JSCS} and the references therein). On the one hand, CRSN is required to
aggregate application-specific data with limited energy. On the
other hand, CRSN nodes should restrict the interference to Primary
User (PU) systems with their intrinsic spectrum sensing capability.
As a smart combination of Cognitive Radio Networks (CRNs) and WSNs,
CRSN has yielded many open research issues which are distinct from
existing ones. Among them, how to design energy efficient
spectrum-aware clustering schemes, in order to effectively organize
and maintain such a large-scale sensor network in a dynamic radio
environment, remains a big challenge.

While much attention has been paid to the clustering issue in either
WSNs or CRNs, few of these works are fully applicable to CRSN. Existing
cognitive radio clustering schemes aim to facilitate joint spectrum and
routing decisions, but seldom jointly consider 1) CRSN's main objective: fast and accurate
acquisition of application-specific source information; 2) CRSN's
additional resource constraint: the energy scarcity problem inherited from
traditional WSNs. The studies in \cite{cluster-CogMesh} and
\cite{cluster-CRN1} seek to minimize the number of clusters in
cognitive mesh networks while ensuring the connectivity of all nodes.
The author of \cite{cluster-CRN3} investigates the route discovery
and repair strategies for clustered CRN. The above mentioned
clustered structures aim at guaranteeing network connectivity under a
dynamic spectrum environment. However, none of them is designed for
the purpose of efficient source information aggregation under energy
constraints.

Similarly, clustering schemes for non-cognitive WSNs are designed
with the main objective of collecting source information with
minimized power consumption. However, they cannot deal with the
spectrum-aware sensing and communication in a cognitive radio
context. In \cite{LEACH}, an energy efficient LEACH protocol is
proposed, where the cluster heads are selected with predetermined
probability, and then other nodes join their specific nearest
cluster heads. Another approach named `HEED' is developed in
\cite{HEED} for clustering ad hoc sensor networks, which chooses the
sensor nodes with more neighboring nodes and larger residual energy
as cluster heads through coordinated election. These algorithms
manage to prolong the network lifetime. However, all of them assume
fixed channel allocation and none can handle dynamic spectrum
access, and thus are not suitable for CRSN.

To accommodate CRSN's unique features, we model communication
power consumption and derive the optimal number of clusters in CRSN.
We prove that minimizing the communication power is equivalent to
minimizing the sum of squared distance between CRSN nodes and their
cluster centers. This objective coincides with many clustering
problems \cite{Clustering-Analysis1}\cite{Clustering-Analysis2}, and the
ideas of constrained clustering \cite{Constrained-K-means}\cite{Constrained-Agglomerative} can be
employed to cluster CRSN nodes under spectrum-aware constraints. We propose a novel distributed
spectrum-aware clustering (DSAC) protocol to form clusters with low
intra-cluster distance and hence reduces communication power.
Moreover, DSAC is performed in a fully self-organized way, and has
preferable scalability and stability.

This paper is outlined as follows. In section II, we introduce
a spectrum-aware clustered structure and model the communication power
consumption model for CRSN. The energy efficient spectrum-aware
clustering schemes are proposed in section III. In section IV, performance evaluation in terms of energy consumption, scalability
and stability is given. Finally, conclusions are
drawn in Section V.

\section{Energy Consumption Model for Cognitive Radio Sensor Network}

\subsection{Spectrum-Aware Clustered Structure}
The differences between our proposed clustered structure from existing
ones are twofold. On the one hand, unlike most clustered topologies
for non-cognitive WSNs, it is aware of the radio environment. To restrict the interference to PUs, only short
distance communications are allowed, by the way of intra-cluster
aggregation and inter-cluster relaying. On the other hand, this
structure should consider the energy saving issue in intra-cluster data
aggregation and inter-cluster relaying. Therefore, our clustered structure is different
from the clustered structure designed for most CRNs, which mainly
consider the channel availability and network connectivity while
putting away the energy issue.

In addition to the aforementioned features, the following basic
assumptions and objectives are used in this paper:

\begin{itemize}
\item Spectrum Sensing Capability: Equipped with spectrum sensing capability, each CRSN node can correctly determine the available channels at its location.
\item Spectrum-Aware Constraint: CRSN nodes that belong to the same cluster should have at least one common available channel, which is not occupied by neighboring PU nodes for the moment.
\item Efficient Application-oriented Source Sensing: We put a cluster head (CH) in every cluster. The sensed source information should be first aggregated to CH, and then relayed to the sink node.
\item Energy Saving Objective: The clusters are organized such that the total communication power is minimized, in order to extend the lifetime of the CRSN.
\end{itemize}

\begin{figure}[t] \centering
\includegraphics[width=0.5\textwidth]{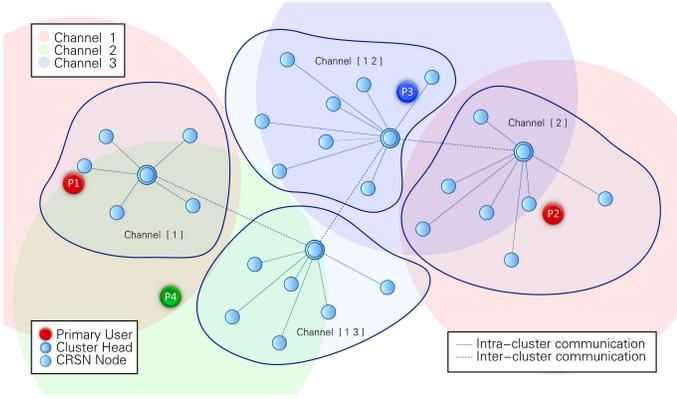}
\caption {An example of spectrum-aware clustered structure for CRSN}
\label{Example_Spectrum-Aware_Clustering}
\end{figure}

The proposed spectrum aware clustering structure is depicted in
Fig.~\ref{Example_Spectrum-Aware_Clustering}. PUs occupying
different channels are represented in corresponding colors. These
channels are not available to CRSN nodes located within the PU's
protected range (translucent area). Neighboring nodes who share
common channels form a cluster and one node has to be selected as CH
in each cluster. The network communication can be categorized into
two classes: intra-cluster communication and inter-cluster
communication. During intra-cluster communication phase, all the
CRSN nodes send their readings of source information to their CH
through the local common channel. During the inter-cluster communication
phase, the CH first compress the aggregated source information, then
transmit it to the upstream neighbor CH using maximal power. With
this structure, the sensed source information is collected
efficiently through intra-cluster aggregation and inter-cluster
relaying.

\subsection{Minimizing Communication Energy}
CRSN nodes inherit the energy constraint from traditional WSNs.
Therefore, how to properly model and minimize the network-wide
communication power becomes our major challenge. We assume that
there are $N$ CRSN nodes and $K$ clusters. The $k$th cluster is
denoted as $c_k$ and has $N_k$ CRSN nodes. The $i$th node of $c_k$
is $n_i^k$, whose coordinate is $\left( {x_i ,y_i } \right)$.

As mentioned before, the power consumption consists of two parts:
inter-cluster power communication and intra-cluster communication
power. Since all these communications are confined within short
distances, free space channel model is applied with $d^2$ power
loss.

In inter-cluster communication stage, the CH compresses and forwards
the collected source information to the sink node through the vacant
channels shared with upstream clusters. The inter-cluster power is
fixed at maximum to improve network connectivity. The sum power for
inter-cluster communication can be expressed as:

\begin{equation}\label{inter-cluster-power}
P_{{\rm{inter}}}  = \sum \limits_{k = 1}^K {P_{{\rm{IC}}}} = K C_0
P_r d_{{\rm{max}}}^2
\end{equation}
where $C_0$ is a loss factor related constant, $P_r$ is the minimal
receiving power required for successful decoding, and
$d_{{\rm{max}}} $ is the maximal transmission range of CRSN node.

Since CH consumes more power than other CRSN nodes, if we fix
certain node as CH, its energy will deplete sooner than other nodes.
To balance the energy consumption within a cluster, we adopt a CH
rotation strategy, which allows all CRSN nodes to take equal
probability to become CH.

\begin{figure}[h] \centering
\includegraphics[width=0.3\textwidth]{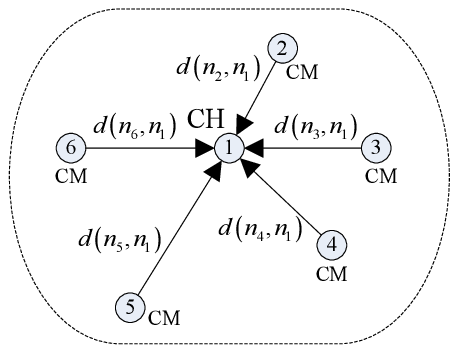}
\caption {Intra-Cluster Communication}
\label{Intra-Cluster_Communication}
\end{figure}

When the $j$th node is selected as CH, all other Cluster Members
(CM) report to CH, as shown in
Fig.~\ref{Intra-Cluster_Communication}. The sum intra-cluster power
is:
\begin{equation}
P_{{\rm{intra}}} \left( {\rm{CH}} = n_j^k  \right) =
\sum\limits_{\scriptstyle i = 1 \hfill \atop
  \scriptstyle i \ne j \hfill}^{N_k } {P_t \left( {n_i^k } \right)}  = C_0 P_r \sum\limits_{\scriptstyle i = 1 \hfill \atop
  \scriptstyle i \ne j \hfill}^{N_k } {d^2 \left( {n_i^k ,n_j^k } \right)}
\end{equation}
where $P_r$ is the minimal receiving power required, and $d\left(
{n_i^k ,n_j^k } \right)$ is the Euclidean distance between the $i$th
and $j$th node which can be acquired through channel estimation.

Taking into account the fact that all CRSN nodes are equally likely
to become CH, the average network-wide intra-cluster communication
power can be estimated as:
\begin{equation}\label{intra-cluster-power}
\begin{array}{l}
 P_{{\rm{intra}}} = \sum\limits_{k = 1}^K {C_0  \cdot P_r \sum\limits_{j = 1}^{N_k } {\frac{1}{{N_k }}} \sum\limits_{\scriptstyle i = 1 \hfill \atop
  \scriptstyle i \ne j \hfill}^{N_k } {d^2 \left( {n_i^k ,n_j^k } \right)} }  \\
  = 2C_0  \cdot P_r \sum\limits_{k = 1}^K {\sum\limits_{i = 1}^{N_k } {d^2 \left( {n_i^k ,center\left( k \right)} \right)} }  \\
 \end{array}
\end{equation}
where $center\left( k \right) = \left( \frac{1}{N_k} \sum\limits_{i
= 1}^{N_k } {x_i^k } ,\frac{1}{N_k} \sum\limits_{i = 1}^{N_k }
{y_i^k } \right)$ is the center of the $k$th cluster.

\subsection{Optimal Number of Clusters}
Now we have $N$ CRSN nodes, and we wish to partition them into $K$
clusters. How many clusters should be created is critical in our
energy saving issue. For instance, if $K=N$ i.e. each CRSN node is
an independent cluster, and all CRSN nodes act as CHs and have to
transmit using maximal power. On the contrary, if $K=1$ i.e. all $N$
CRSN nodes form a single cluster, the intra-cluster communication
energy will be too high due to very far intra-cluster distances.
Both of these two extreme cases will result in excessive energy
consumption. As a result, optimal number of clusters should be
carefully chosen to effectively save network-wide energy. For
uniformly distributed CRSN nodes, we analytically derive the optimal
number of clusters that can statistically minimize the network-wide
energy consumption.

From \eqref{inter-cluster-power} and \eqref{intra-cluster-power},
the expectation of total communication power is:
\begin{equation}\label{E_P_network1}
\begin{array}{l}
 P_{{\rm{total}}}  =  \\
 2C_0 P_r \cdot E\left( {\sum\limits_{k = 1}^K {\sum\limits_{i = 1}^{N_k } {d^2 \left( {n_i^k ,center\left( k \right)} \right)} } } \right) + K\cdot C_0 P_r \cdot d_{{\rm{max}}}^2  \\
 \end{array}
\end{equation}

It is reasonable to assume that the randomly deployed CRSN nodes are
uniformly distributed in the 2-dimensional area around the center
point, and the density $\rho$ is predetermined by
application-specific source sensing mission. Therefore:

\begin{equation}
\begin{array}{l}
 E\left( {\sum\limits_{k = 1}^K {\sum\limits_{i = 1}^{N_k } {d^2 \left( {n_i^k ,center\left( k \right)} \right)} } } \right) \\
  = N\left( {{\mathop{\rm var}} \left( {x_i^k } \right) + {\mathop{\rm var}} \left( {y_i^k } \right)} \right) = \frac{{Nd^2 }}{6} \\
 \end{array}
\end{equation}
, where $d$ is the average diameter of a cluster.

Since there are $\frac{N}{K}$ nodes per cluster on average, and the
density of CRSN nodes is $\rho$. Then, the area of the cluster can
be estimated as $d^2 = \frac{N}{{K\rho }}$.

Substituting the above formulations into \eqref{E_P_network1}, we
get:

\begin{equation}\label{E_P_network2}
E\left( {P_{{\rm{total}}} } \right) = C_0 P_r \left( {\frac{{N^2
}}{{3\rho K}} + Kd_{{\rm{max}}}^2 } \right)
\end{equation}

Obviously, \eqref{E_P_network2} is a convex function and the optimal
number of clusters can be estimated by setting its derivative with
respect to $K$ to zero. The optimal number of clusters should be
rounded to an integer:

\begin{equation}
K_{{\rm{opt}}}  = \left\lfloor {\frac{N}{{d_{{\rm{max}}} \sqrt
{3\rho } }} + 0.5} \right\rfloor
\end{equation}

\section{Energy Efficient Spectrum-Aware Clustering}

\subsection{Groupwise Constrained Agglomerative Clustering}
After the optimal number of clusters $K_{{\rm{opt}}}$ is determined,
the communication power is only influenced by intra-cluster part.
Hence, according to \eqref{intra-cluster-power}, minimizing
communication power is equivalent to minimizing sum of squared
distance between CRSN nodes and their cluster centers:

\begin{equation}\label{Sum_Square_Distance}
{\rm{mininize}} \ P_{{\rm{total}}}  \Leftrightarrow {\rm{mininize}}
\ \sum\limits_{k = 1}^K {\sum\limits_{i = 1}^{N_k } {d^2 \left(
{n_i^k ,center\left( k \right)} \right)} }
\end{equation}

In clustering analysis theory
\cite{Clustering-Analysis1}\cite{Clustering-Analysis2},
\eqref{Sum_Square_Distance} is called sum of squared error (SSE),
also known as `scatter'. Minimizing SSE is also the goal of many
clustering algorithms. Therefore, the energy saving objective
coincides with many clustering analysis problems and we can employ
the ideas in clustering analysis theory to design desirable
clustering schemes. Some computationally feasible heuristic
clustering methods have been well developed. The main techniques
include K-means, Fuzzy C-means, and Hierarchical Clustering, etc.
Several of them are effective in clustering non-cognitive WSNs
\cite{FCM-WSN}.

However, CRSN nodes should have at least one common available
channel to form a cluster. These requirements are imposed on the
clustering problem as spectrum-aware constraints, as expressed in
\eqref{Spectrum-Aware_Constraint}. Therefore, the existing
clustering schemes in non-cognitive WSNs are inapplicable for CRSN,
since all of these algorithms do not consider the spectrum-aware
constraint.

\begin{equation}\label{Spectrum-Aware_Constraint}
\left| {Chan\left( {n_i^k } \right) \cap Chan\left( {n_j^k }
\right)} \right| \ge 1,n_i^k ,n_j^k  \in Cluster\left( k \right)
\end{equation}
where $\left| {Chan\left( {n_i^k } \right)} \right|$ denotes the
number of available channels for $n_i^k$.

In recent years, a branch of constrained clustering algorithms have
been developed to cluster instances with pairwise constraints, such
as constrained K-means \cite{Constrained-K-means} and constrained
complete-link clustering \cite{Constrained-Agglomerative}. Pairwise
constraints are imposed on pairs of nodes to influence the outcome
of clustering algorithm and they mainly include two types: must-link
and cannot-link constraints.

\begin{figure}[h] \centering
\includegraphics[width=0.45\textwidth]{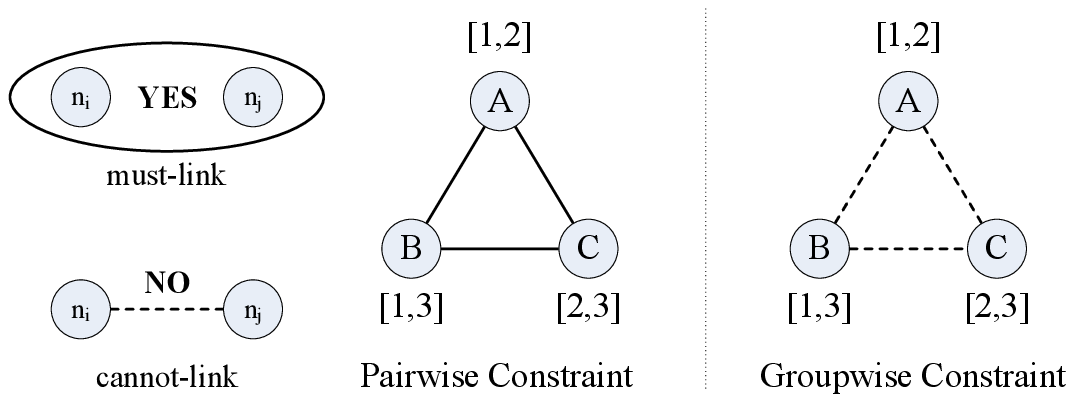}
\caption {Comparison of Pairwise Constraint and Groupwise
Constraint} \label{Pairwise-Groupwise}
\end{figure}

As shown in Fig.~\ref{Pairwise-Groupwise}, The must-link constraint
forces $n_i$ and $n_j$ to be in the same cluster, while the
cannot-link constraint specifies that $n_i$ and $n_j$ must not be
placed the same cluster. If two CRSN nodes have no available
channels in common, they can not be allocated into one cluster, and
this is equivalent to imposing cannot-link constraint on this node
pair. Thus, the ideas of constrained clustering algorithms can be
used to design spectrum-aware clustering scheme for CRSN. However,
existing constrained clustering methods can not be directly applied
to the spectrum-aware clustering, since our spectrum-aware
constraints are imposed on groups, rather than pairs.

Now, we define `groupwise constraint' by explaining the differences
between `pairwise constraint' and `groupwise constraint'. In
Fig.~\ref{Pairwise-Groupwise}, three nodes can operate on three
channels, and the numbers labeled beside the nodes represent the
available channels. On the middle and right, node A and B share
channel 1, A and C share channel 2, and B and C share channel 3. If
employing pairwise constraint, each node pair shares a common
channel and no `cannot-link constraint' is imposed on them, then
they can form one cluster. However, if groupwise constraint is
imposed, the three nodes share no common channel and can not form a
cluster.

The spectrum-aware constraint is a kind of groupwise constraint. All
existing pairwise constrained clustering algorithms are iterative,
and the basic idea is to satisfy the pairwise constraints in each
single iteration. In order to extend the existing algorithms to the
model with groupwise constraint, we have to replace pairwise
constraint with spectrum-aware groupwise constraint. Here, we impose
the spectrum-aware constraint on the complete-link agglomerative
clustering algorithm \cite{Constrained-Agglomerative} to cluster the
CRSN, and name it the `Groupwise Constrained Agglomerative
Clustering' (GCAC). The basic idea of GCAC is to set each node as a
disjoint cluster at the beginning and then merges two nearest
clusters in each iteration until the cluster number reduce to the
optimal number. In each iteration, the inter-cluster distances
should be re-calculated according to complete-link principle.

\subsection{Distributed Spectrum-Aware Clustering}
Although GCAC can produce clusters satisfying spectrum-aware
constraints, it requires some central processor with global node
information to perform the clustering algorithm. This is impractical
and conflicts with the distributed nature of CRSN. To address this
problem, we propose a novel Distributed Spectrum-Aware Clustering
(DSAC) protocol which can form clusters in a fully self-organized
fashion. The basic idea of DSAC inherits that of GCAC in general:
the closest nodes with common channel will agglomerate into a small
group first and then the other neighboring nodes will join in one
after another. The main differences are as follows: GCAC compares
the distance between all clusters and find the global minimum pair
to merge first, while DSAC only needs the local minimum distance
through neighborhood information exchange and merges the locally
closest pair.

DSAC protocol is described by the pseudocode shown in
\textbf{Algorithm~\ref{DSAC}}. It consists of three stages: channel
sensing, beaconing and coordination. In channel sensing stage, every
CRSN node determines the vacant channels individually and compares
it with the previously sensed result. In beaconing stage, CRSN node
beacons its node information on the detected vacant channels. If any
PU state change is detected, the node declares itself as a new
cluster by beaconing a new cluster ID. Otherwise, the node stays
with the current cluster. After the node beaconing, the CH updates
and beacons the cluster information, including cluster size and
common channels. In intra-cluster coordination stage, each node in a
cluster first measures the strength of neighboring beacon signals
and then announces the pairwise distances. Then, CH determines the
inter-cluster distance according to groupwise constraint and
complete-link rule \cite{Clustering-Analysis2}, in which the
distance is jointly decided by the common available channels and the
max distance between the nodes of two clusters. In inter-cluster
coordination, every CH will send a merge invitation to its nearest
neighbor cluster. If any two clusters send merge invitations to each
other, they merge into a single cluster by unifying new cluster ID
and common channels and selecting a new CH. Otherwise, the cluster
just needs to select a new CH while the topology remaining
unchanged.

\begin{algorithm}
\caption{Distributed Spectrum-Aware Clustering} \label{DSAC}
\begin{algorithmic}
\STATE \textbf{Initialize ()} \STATE Define every node as a disjoint
cluster: $c_i \left( {n_i } \right) \Leftarrow n_i$ \STATE
\textbf{Channel Sensing ()} \STATE $Chan\left( {n_i } \right),i \in
1, \cdots ,N$ \STATE \textbf{Distributed Spectrum-Aware Clustering
()} \STATE 1. Node Beacon \FOR{$i \in 1, \cdots ,N$} \STATE Beacon
Node ID \IF{$Chan^t \left( {n_i } \right) = Chan^{t - 1} \left( {n_i
} \right)$} \STATE Beacon Cluster ID: $c_i^t \left( {n_i } \right) =
c_i^{t - 1} \left( {n_i } \right)$ \ELSE \STATE Beacon New Cluster
ID: $c_i^t \left( {n_i } \right) = {\rm{new \ CID}}$ \ENDIF \ENDFOR
\STATE 2. Cluster Beacon: \FOR{$m \in 1, \cdots ,K$} \STATE Beacon
ClusterSize: $\left| {c_m^t } \right|$, CommonChannel: $Chan^t
\left( {c_m } \right)$ \ENDFOR \STATE 3. Intra-cluster Coordinate:
\FOR{$m \in 1, \cdots ,K$} \IF{received node beacon number ${\rm{
}}c_n ^{\rm{r}} {\rm{ < }}\left| {c_n^t } \right|$} \STATE $d\left(
{c_m ,c_n } \right) \Leftarrow \infty$ \ELSE \STATE $d\left( {c_m
,c_n } \right) = \mathop {\max }\limits_{n_i  \in c_m ,n_j  \in c_n
} d\left( {n_i ,n_j } \right)$ \ENDIF \STATE Find nearest neighbor
cluster: \STATE $n = \mathop {\arg \min }\limits_{n \in
Neighbor\left( m \right)} d\left( {c_m ,c_n } \right)$ \ENDFOR
\STATE 4. Inter-cluster Coordinate: $c_m$ send merge invitation to
$c_n$ \IF{$c_m$ also receive merge invitation from $c_n$} \STATE
Unify New Cluster ID: $c_l  \Leftarrow \left( {c_m ,c_n } \right)$
\STATE Assign Channel: $Chan\left( {c_l } \right) = \left|
{Chan\left( {c_m } \right) \cap Chan\left( {c_n } \right)} \right|$
\ENDIF \STATE 5. Merge Complete: Select new CH \STATE goto
\textbf{Channel Sensing ()}
\end{algorithmic}
\end{algorithm}

\begin{figure}[h] \centering
\includegraphics[width=0.45\textwidth]{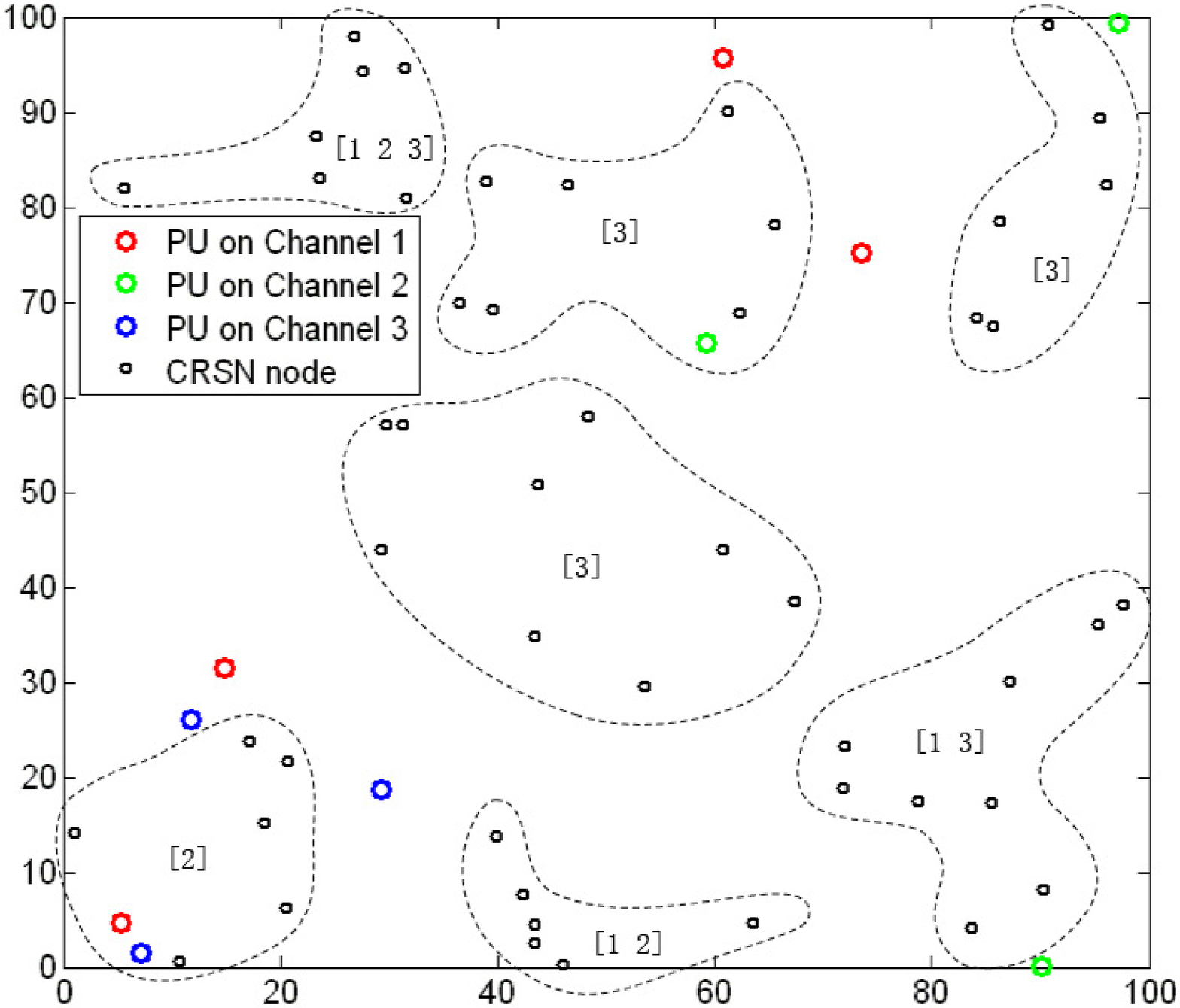}
\caption {An example of the DSAC clustering result}
\label{Example_DSAC}
\end{figure}

Fig.~\ref{Example_DSAC} shows an example of the DSAC clustering
result, where 50 CRSN nodes and 10 PUs are randomly deployed on a
100 meter $\times$100 meter field. There are three available
channels in the system (marked by red, green and blue). The
clustering result is illustrated by dashed enclosures and the
corresponding common channels are labeled in the cluster.

\section{Performance Evaluation}
In this section, we analyze and simulate the performance in terms of
scalability, energy consumption and stability. We employ Monte Carlo
experiments and repeat a hundred thousand times to compute the
target value.

In order to evaluate the performance of the proposed DSAC scheme, we
have to employ a generally accepted algorithm called K-means
clustering as a reference. According to literature, K-means can
converge to local minimal SSE in very short time. Although K-means
does not include the spectrum-aware constraint and is only
applicable for non-cognitive WSNs, it serve as a good criterion for
performance evaluation.

For all the experiments, we randomly deploy PUs and CRSN nodes in a
$100 \times 100$ square meters area. The PUs can operate on three
channels, and CRSN nodes can only access the channels on which the
neighboring PUs are inactive. Every PU randomly occupies one of the
three channels. The protection range for PU is 20 meters, which
means the PU's CRSN neighbors within this range can not access its
occupied channel.

For GCAC algorithm, the time complexity is similar to the existing
complete-link agglomerative clustering algorithms, which is $O(N^2
\log N)$\cite{Clustering-Analysis2}. Although this complexity is
much lower than the exhaustive method and can be well implemented in
some small sensor networks, it is still too high to be implemented
in the large scale CRSN. Relatively, the time complexity of K-means
is much lower, which is $O(N¡ÁK)$.

\begin{figure}[h] \centering
\includegraphics[width=0.45\textwidth]{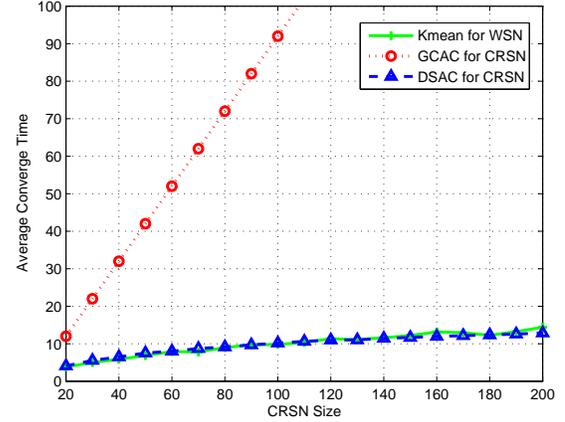}
\caption {Scalability: CRSN size vs converge time}
\label{Compare_CRSNsize_ConvergeTime}
\end{figure}

In the first experiment, we simulate the average converge time of
the three clustering schemes when CRSN size is growing. As shown in
Fig.~\ref{Compare_CRSNsize_ConvergeTime}, the converging time of
GCAC grows proportionally with the CRSN size, while the DSAC
converges almost as fast as the efficient K-mean algorithm. This
result shows DSAC has similar time complexity to K-means and
therefore exhibits satisfying scalability.

\begin{figure}[h] \centering
\includegraphics[width=0.45\textwidth]{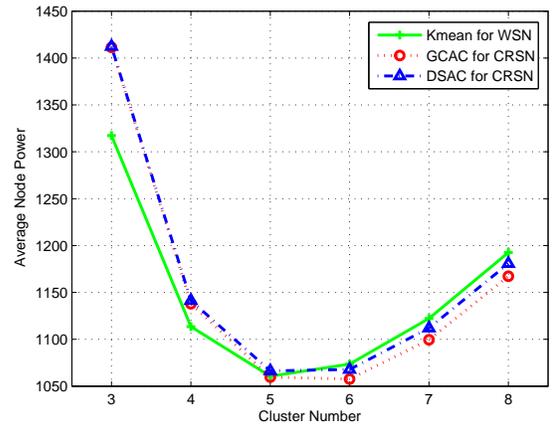}
\caption {Comparison of average energy among three schemes}
\label{Compare_Average_Energy}
\end{figure}

For the following experiments, we assume the max transmission range
for CRSN node is 50 meters, and 20 CRSN nodes and 5 PU nodes are
uniformly distributed in the same area. According to the theoretical
analysis in Section II.C, the estimated optimal cluster number is
about five. In the simulation, we set the cluster number from 3 to
8, and calculate the average power consumed by CRSN nodes. From
Fig.~\ref{Compare_Average_Energy}, we find that the minimum power
occurs when cluster number is about 5 to 6, and this result agrees
well with the theoretical analysis.

\begin{figure}[h] \centering
\includegraphics[width=0.45\textwidth]{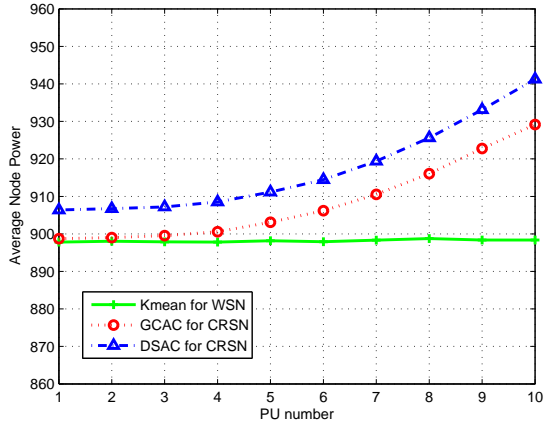}
\caption {PU number vs average node power}
\label{Compare_nPU_NodePower}
\end{figure}

To evaluate the influence of PUs on clustering, we simulate the
average CRSN node power consumption when different numbers of PU
node are active. In Fig.~\ref{Compare_nPU_NodePower}, we set the
CRSN size as 30 and adjust the PU number from 1 to 10. For
non-cognitive WSN, the K-means clustering result is not influenced
by PU systems, therefore the average node power keeps steady. For
CRSN, as more PU nodes are active, more spectrum-aware constraints
are imposed on the clustering process. Therefore the clustering
results are poorer in terms of energy consumption. Again, we find
the performance of DSAC only to be slightly worse than that of GCAC.

\begin{figure}[h] \centering
\includegraphics[width=0.45\textwidth]{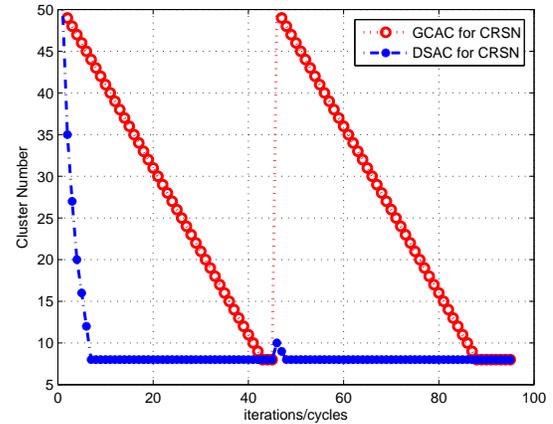}
\caption {Comparison of dynamic stability between GCAC and DSAC}
\label{Compare_Dynamic_Stability}
\end{figure}

In the final experiment, we examine the proposed algorithm's
stability under dynamic PU activities. For exhaustive search method,
K-means and GCAC, if any PU activity or CRSN node position changed,
the whole network should be involved in re-clustering, which makes
the network topology less stable and requires extra control
overhead. However, in DSAC, only the nodes that detect PU activity
change will engage in re-clustering. In
Fig.~\ref{Compare_Dynamic_Stability}, when one PU changes its
status, only 3 of 50 CRSN nodes are affected. After two merges, the
network once again converges to stable topology, which is much
faster than GCAC. During the re-clustering, the rest nodes' status
and their clustering structure remain the same. Their
application-specific sensing task won't be influenced. Hence, the
stability of network is preserved as much as possible.

\section{Conclusion}
In this paper, we proposed a novel distributed spectrum-aware
clustering scheme for cognitive radio sensor networks. We modeled
the communication power for CRSN, which consists of intra-cluster
aggregation and inter-cluster relaying. After deriving the optimal
number of clusters, we minimize the CRSN energy using groupwise
constrained clustering, in which the spectrum-aware requirement is
regarded as groupwise constraint. With the proposed DSAC protocol,
desirable clustering results can be produced. Through extensive
simulations, we find that DSAC has preferable scalability and
stability because of its low complexity and quick convergence under
dynamic PU activity change.

%\section{Acknowledgement}
%This work was supported in part by the National Science and
%Technology Major Project of China (Nos. 2009ZX03003-004-03 and
%2010ZX03003-003-01), the National Key Basic Research Program of
%China (No. 2009CB320405), the Natural Science Foundation of China
%(No. 60972057), the Program for New Century Excellent Talents in
%University (NCET-09-0701), the Fundamental Research Funds for the
%Central Universities of China, and the US National Science
%Foundation under Grant CCF-0830462 and ECCS-1002258.

\end{document}